\documentclass{optica-article}
\journal{oe}

\usepackage{indentfirst}

\usepackage{fancyhdr} 
\usepackage[section]{placeins}
\begin{document}

%\preprint{BAQIS - Quantum Photonics}

\title{Ultra-narrowband interference circuits\\ enable low-noise and high-rate photon counting for InGaAs/InP avalanche photodiodes}

\author{Yuanbin~Fan,\authormark{1,\dag} Tingting~Shi,\authormark{1,2,3,\dag} Weijie~Ji,\authormark{1}Lai~Zhou,\authormark{1} Yang~Ji \authormark{2,3} and Zhiliang~Yuan\authormark{1,*}}

\address{\authormark{1}Beijing Academy of Quantum Information Sciences, Beijing 100193, China\\
\authormark{2}State Key Laboratory for Superlattices and Microstructures, Institute of Semiconductors, Chinese Academy of Sciences, Beijing 100083, China\\
\authormark{3}College of Materials Science and Opto-Electronic Technology, University of Chinese Academy of Sciences, Beijing 100049, China\\
\authormark{\dag}These authors contributed equally.}

\email{\authormark{*}yuanzl@baqis.ac.cn} %% email address is required; see note below about the corresponding author designation



\begin{abstract}
Afterpulsing noise in InGaAs/InP single photon avalanche photodiodes (APDs) is caused by carrier trapping and can be suppressed successfully through limiting the avalanche charge via sub-nanosecond gating. Detection of faint avalanches requires an electronic circuit that is able to effectively remove the gate-induced capacitive response while keeping photon signals intact. 
Here we demonstrate a novel ultra-narrowband interference circuit (UNIC) that can reject the capacitive response by up to 80~dB per stage with little distortion to avalanche signals.  Cascading two UNIC's in a readout circuit,  we were able to enable a
high count rate of up to 700~MC/s and a low afterpulsing of 0.5~\% at a detection efficiency of 25.3~\%
for 1.25~GHz sinusoidally gated InGaAs/InP APDs.  At a temperature of -30~$^\circ$C, we measured an afterpulsing probability of 1~\% at a detection efficiency of 21.2~\%.  
\end{abstract}

%\keywords{Single Photon Avalanche Diode (APD), afterpulsing,}

%\clearpage

\section*{Introduction}

Semiconductor avalanche photodiodes (APD's) are versatile for weak light detection, with applications from remote ranging\cite{Wehr1999,Schreiber1999}, quantum communication \cite{yuan2018} and fluorescence lifetime imaging\cite{Damalakiene2016} to optical time-domain reflectometry \cite{Healey1984,Eraerds2010}.
For practical fiber quantum key distribution (QKD), InGaAs/InP APD's are the detector of choice because they are compact and low cost, and allow cryogenic-free or even room-temperature operation \cite{yuan2018}.
However, they suffer from spurious afterpulsing arising from carrier trapping by defects in the multiplication layer, especially at high detection efficiencies \cite{Comandar2015,Tada2020}.   
To minimise afterpulsing,  an APD can be biased on for a sub-nanosecond duration only when a photon arrival is expected. 
In doing so, charge per avalanche can be reduced to the order of 10~fC \cite{yuan2010,Restelli2013,Namekata2006}, corresponding to a transient current of less than 0.1~mA. 
Such weak avalanches have to be discriminated through use of a readout circuit that removes the strong capacitive response to the applied gates. Presently, gated InGaAs detectors are capable of counting photons at up to 60\% efficiencies \cite{Fang2020} and 1~GHz rate \cite{patel2012} and with photon number resolution \cite{kardynal2008}.  Thanks to this success, gating approach has been applied to traditionally free-running Si devices for performance enhancement \cite{Thomas2010,wayne2021}.   

Existing readout circuits include band stop \cite{Namekata2006,Namekata09,Tada2020} or low-pass\cite{walenta2012,He2017,Fang2020} filtering under sine-wave gating\cite{Namekata2006}, self-differencing \cite{Yuan2007,Comandar2015}, and transient reference cancellation \cite{Restelli2013,Liang2019}. While simple for implementation, frequency filtering \cite{Namekata2006,Namekata09,Tada2020,walenta2012,He2017,Fang2020} distorts the avalanche signals due to its rejection of a sizeable portion of frequency components, thus increasing time jitter and temporal errors in photon registrations \cite{walenta2012}.  Self-differencing \cite{Yuan2007} and reference cancellation methods \cite{Restelli2013} are able to maintain avalanche signal fidelity but may suffer operational complexities. 
The former requires a wideband performance for the entire circuitry and thus inconveniently an adjustable delayline \cite{yuan2010} for frequency alignment, while the latter \cite{Restelli2013} can be unstable because the transient reference is derived separately from the capacitive response. 

Here we propose and experimentally demonstrate a simple, low-distortion ultra-narrowband interference circuit (UNIC)
that can suppress the capacitive response for a 1.25~GHz gated InGaAs/InP APD single photon detector.
The circuit is an asymmetric radio-frequency (RF) interferometer. One of its arms contains a narrow band pass filter (BPF) based on surface acoustic wave resonator (SAW) to retrieve the fundamental wave of the gating signal.
The filtered wave then interferes destructively with the same frequency component transmitted via the other arm through a coupling module, thereby eliminating the capacitive response. 
This interference occurs over a narrow band, so it can provide a broad 
and continuous pass band in frequency domain to maintain the avalanche signal with little distortion. 
This allows to achieve ultra-low afterpulsing probabilities and an excellent jitter performance at high detection efficiencies from two InGaAs APD's that exhibit capacitive responses of very different amplitudes.   

\section*{Detector characterisation setup}

\begin{figure*}[htb]
\centering
\includegraphics[width=.6\linewidth]{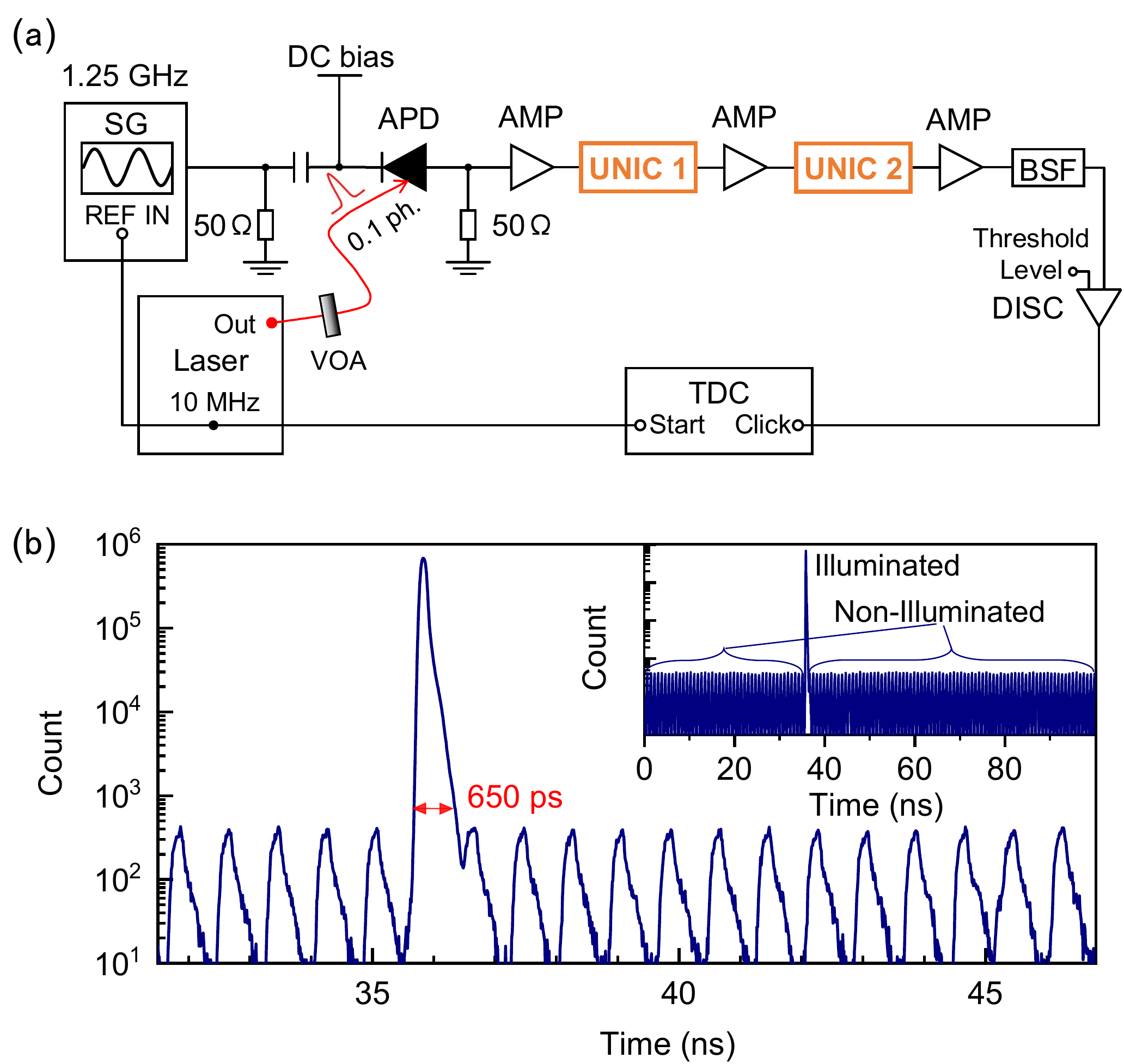}
\caption{
\textbf{(a)} Single-photon characterisation setup for 1.25~GHz sinusoidally gated InGaAs/InP APDs using UNICs for avalanche impulse readout;  
\textbf{(b)} A histogram of the photon detection events measured by the characterisation setup \textbf{(a)} on an InGaAs APD detector that was regulated at a temperature of 30~$^\circ$C. The photon detection peak exhibits a 30~dB width of 650~ps. 
AMP: amplifier; APD: avalanche photodiode; BSF: band stop filter;  DISC: discriminator;
SG: signal generator; TDC: time-to-digital converter; UNIC: ultra-narrowband interference circuit; VOA: variable optical attenuator.  
}
\label{fig:001}
\newpage
\end{figure*}

Figure~\ref{fig:001}(a) shows our single photon characterisation setup for InGaAs APDs. A 1550~nm passively mode-locked laser serves as the light source and provides stable short pulses of 5-10~ps duration at a repetition rate of 10~MHz.
The laser output power is monitored by an optical power meter of $\pm$5 \% uncertainty (EXFO FTB-1750) and its pulse intensity is set by a variable optical attenuator (VOA, EXFO FTB-3500) to 0.1 photon/pulse at the fiber input of APD under test. It provides a 10~MHz reference to a signal generator (SG) for synthesising a 1.25~GHz sinusoidal wave with up to 27 V voltage swing.  
In combination of a suitable DC bias, this AC signal periodically gates the APD above its breakdown voltage ($60 - 70$~V) to achieve the single photon sensitivity with an effective gate width of 150~ps. 
The APD output is processed by the readout module consisting of two identical 1.25~GHz UNIC's, one 2.5~GHz band stop filter (BSF) of a 10~dB stop band of 100~MHz and three RF amplifiers (AMPs) of 6~GHz bandwidth. Amplification of the raw APD signals is useful as it prevents weak avalanche signals from falling below thermal noise by attenuation of the first UNIC.
The readout signal is discriminated by a discriminator for avalanches before feeding to a time-digital-converter (TDC) with a dead time of 
2~ns for time-resolved photon counting.
Figure~\ref{fig:001}\textbf{(b)} is a typical histogram obtained with this setup.

APD under test is temperature-regulated using their integrated thermal-electric cooler, which is driven by a temperature controller (Thorlabs TED200C). 
A source-measure unit (Keithley 2635B) provides the DC bias and simultaneously monitors the current flowing through the APD.  In characterising the maximum count rate, we replace the 10~MHz laser with a continuous-wave distributed feedback laser (DFB) laser, the output of which is carved into 1.25~GHz, 50~ps pulse train using an intensity modulator. 
We use a high speed digital oscilloscope to record the detector output and extract the count rate through digital discrimination in software. The oscilloscope method is carefully calibrated at low count rate regimes to be consistent with the hardware discriminated result using the photon counter (Stanford Research SR400). 

The setup is able to measure the dark count probability, afterpulsing probability, detection efficiency, maximum count rate, avalanche charge and time jitter. With no performance screening, two fiber-pigtailed APDs from different manufacturers were used in this study, named APD\#1 and APD\#2 respectively.
\section*{Ultra-narrowband interference circuit (UNIC)}

\begin{figure}[ht]
\centering
\includegraphics[width=.8\linewidth]{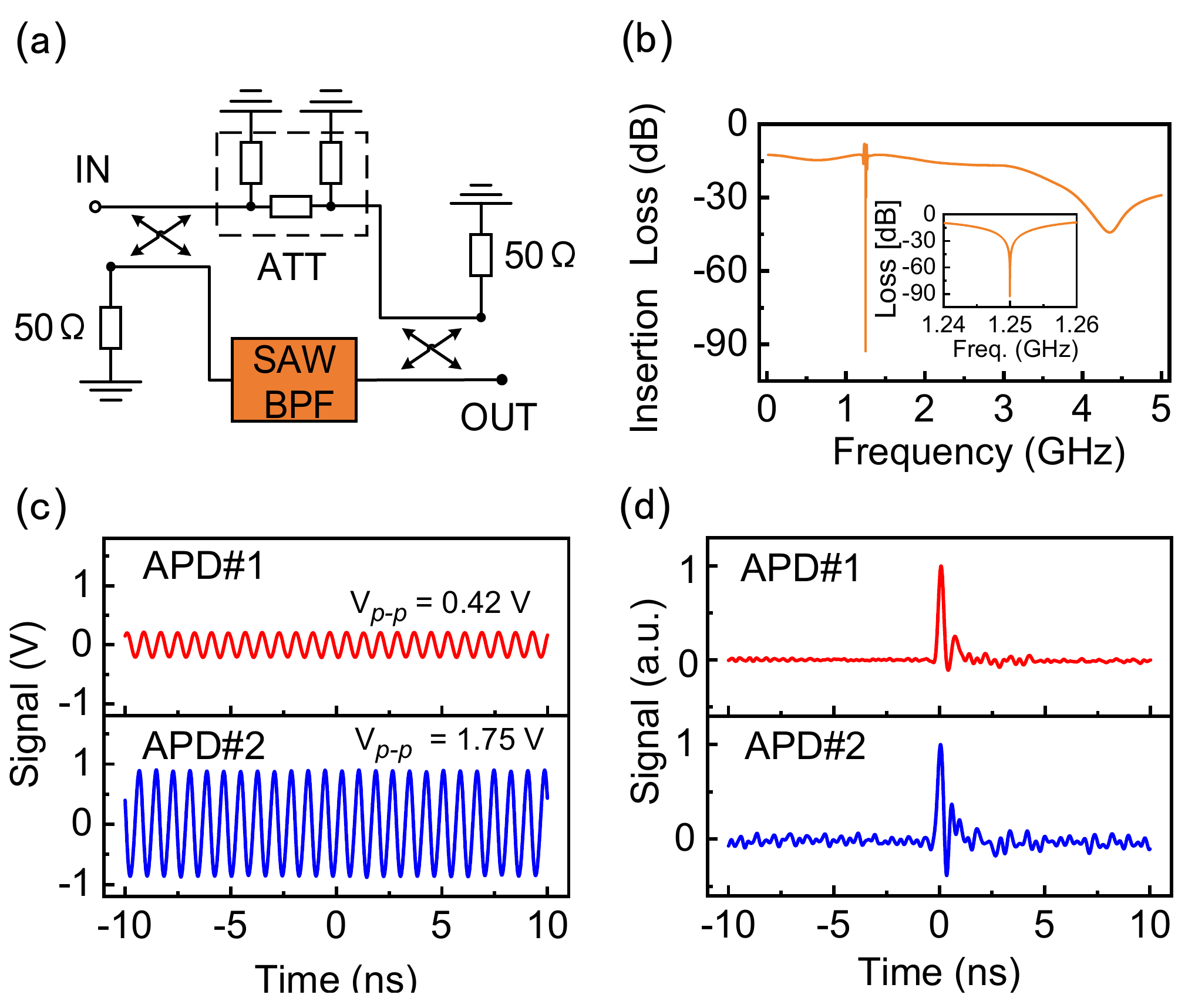}
\caption{\textbf{(a)} Schematic for the ultranarrow interference circuit (UNIC); \textbf{(b)}  Transmission spectrum of a heroic UNIC PCB; Inset: Magnified view for region of 1.24 -- 1.26~GHz. \textbf{(c)} Raw capacitive responses from APD\#1 (top) and APD\#2 (bottom) under identical 27.0~V V$_{p-p}$ gating; \textbf{(d)} Recovered avalanche impulses.  ATT: attenuator; SAW BPF: surface acoustic wave band pass filter.
}
\label{fig:002}
\end{figure}

Under a sub-nanosecond gating,  a photon induced avalanche is an impulse and contains a wide frequency spectrum.  
In contrast, the capacitive response is periodic and has its most energy concentrated at the gating frequency or its higher harmonics. This spectral difference allows a frequency-dependent signal process to remove the capacitive response and keep the wide-band impulses intact. 
Figure~\ref{fig:002}\textbf{(a)} shows a circuit diagram of UNIC. It is an RF interferometer containing two couplers of a 9:1 power splitting ratio, a $\pi$-resistive  attenuator (ATT) and a surface acoustic wave (SAW) band pass filter. 
Two of the ports are terminated by 50~$\Omega$ resistors. 
The SAW BPF features a central frequency of 1.25~GHz, a 20-dB passband of 35~MHz, 
a transmission loss of 3~dB, and a group delay of 34~ns. It filters out the fundamental wave of the gating frequency, which then interferes with the APD signal transmitted through the other arm. 
The attenuation and differential delay are set to enable destructive interference for the 1.25~GHz frequency component at the UNIC output port. The UNIC differential delay ($\Delta t$) meets the condition below 
\begin{equation}
    \Delta t = T_g^{SAW} + \delta t
             = (N+1/2)/f_g,
\end{equation}
\noindent where $T_g^{SAW}$ is the group delay of the SAW BPF, $\delta t$ the delay caused by the track length difference between the two interferometer arms, $f_g = 1.25$~GHz the APD gating frequency, and $N$ is an integer number.  For a compact circuit, we choose $\delta t$ to be less than the half-wave of the gating signal. With the SAW device used, $N = 42$ and $\delta t = 155$~ps. The resulting UNIC unit has a small footprint of $38 \times 15$~mm$^2$ on printed circuit boards (PCBs). 

The large $T_g^{SAW}$ brings two additional benefits. Firstly, it substantially increases the PCB manufacturing tolerance, as a 0.5~mm deviation in the RF track length will just alter the circuit central frequency by less than $10^{-4}$.  This eliminates the requirement of an adjustable delayline \cite{yuan2010} for a precise frequency alignment. Secondly, it helps to produce an ultra-narrow band rejection at its designed frequency.  Figure~\ref{fig:002}(b) shows the measured transmission spectrum  (S21 parameter) of our heroic UNIC PCB, and its inset expands the frequency section of 1.24 -- 1.26~GHz to show the narrowness of the transmission loss dip in the close proximity of the resonance frequency of 1.25~GHz.  The dip of the heroic (typical) PCB features a loss of -95~dB (-80~dB), representing a suppression of 80~dB (65~dB) as compared with the background loss for other frequencies under 2~GHz. 
The dip has a 30~dB linewidth of just 30~kHz, thus ensuring crucial suppression of the APD gating signal without overly distorting the avalanche signals.  
The background loss of about 14~dB is caused mainly by the 9:1 couplers and can be reduced in future with more balanced splitters. 

Cascading two UNIC's enables a stable 100~dB suppression of the primary gating frequency and thus provides a healthy performance redundancy. Their attenuation to the avalanche signal is compensated by using RF amplifiers (Fig.~\ref{fig:001}\textbf{(a)}). Second order harmonics (2.5~GHz) is suppressed by a band stop filter of conventional LC  design. 
Figure~\ref{fig:002}\textbf{(c)} shows raw outputs from two different APD's under identical sinusoidal gating. Their respective capacitive responses are measured to have amplitudes of 0.42~V and 1.75~V.  Despite their 4 times differences,  UNIC's can successfully reject the sinusoidal responses and retrieve avalanches with excellent signal-to-background ratio, as shown in Fig.~\ref{fig:002}\textbf{(d)}.  For APD\#2, we just adjusted the gain of the first AMP to avoid amplification saturation and signal distortion. 

\section*{Results and discussion}

Time-resolved photon counting allows precise extraction of the net photon detection efficiency ($\eta_{net}$) and the afterpulsing probability ($P_A$), which is defined as the ratio of the total afterpulses per photon induced event. 
Figure~\ref{fig:001}\textbf{(b)} shows a histogram of avalanche events measured for APD\#1 under 10~MHz pulsed excitation of 0.1~photon/pulse.   
The illuminated peak has a full-width of 1/1000 maximum (30~dB width) of just 650~ps,  which is shorter than the gating period of 800~ps and thus allows low-error clock number assignment that is essential for high speed QKD. The counts at non-illuminated gates arise from detector dark and afterpulse noise, and their counting rate is 3 orders of magnitude lower than that of the illuminated gate. 
We extract quantities of $P_I$ and $P_{NI}$, \textit{i.e.}, the respective counting probabilities for each illuminated and non-illuminated gate. With a separate measurement of the detector dark count probability ($P_D$), we calculate the afterpulsing probability using the standard method \cite{Yuan2007,Namekata09},
\begin{equation}
	P_A  = \frac{(P_{NI} - P_{D}) \cdot R}{P_I - P_{NI}},
\end{equation}
\noindent where $R = 125$ here is the ratio of the  gating frequency (1.25~GHz) to the laser illumination (10~MHz).
Excluding the dark and afterpulse count probabilities, the net single photon detection efficiency is given by~\cite{Comandar2015}
\begin{equation}
\eta_{net} = \frac{1}{\mu}\mathrm{ln}\frac{1-P_{NI}}{1-P_{I}},
\end{equation}
\noindent where $\mu$ is the average incident photon number per illumination pulse.

\begin{figure}[t]
\centering
\includegraphics[width=.8\linewidth]{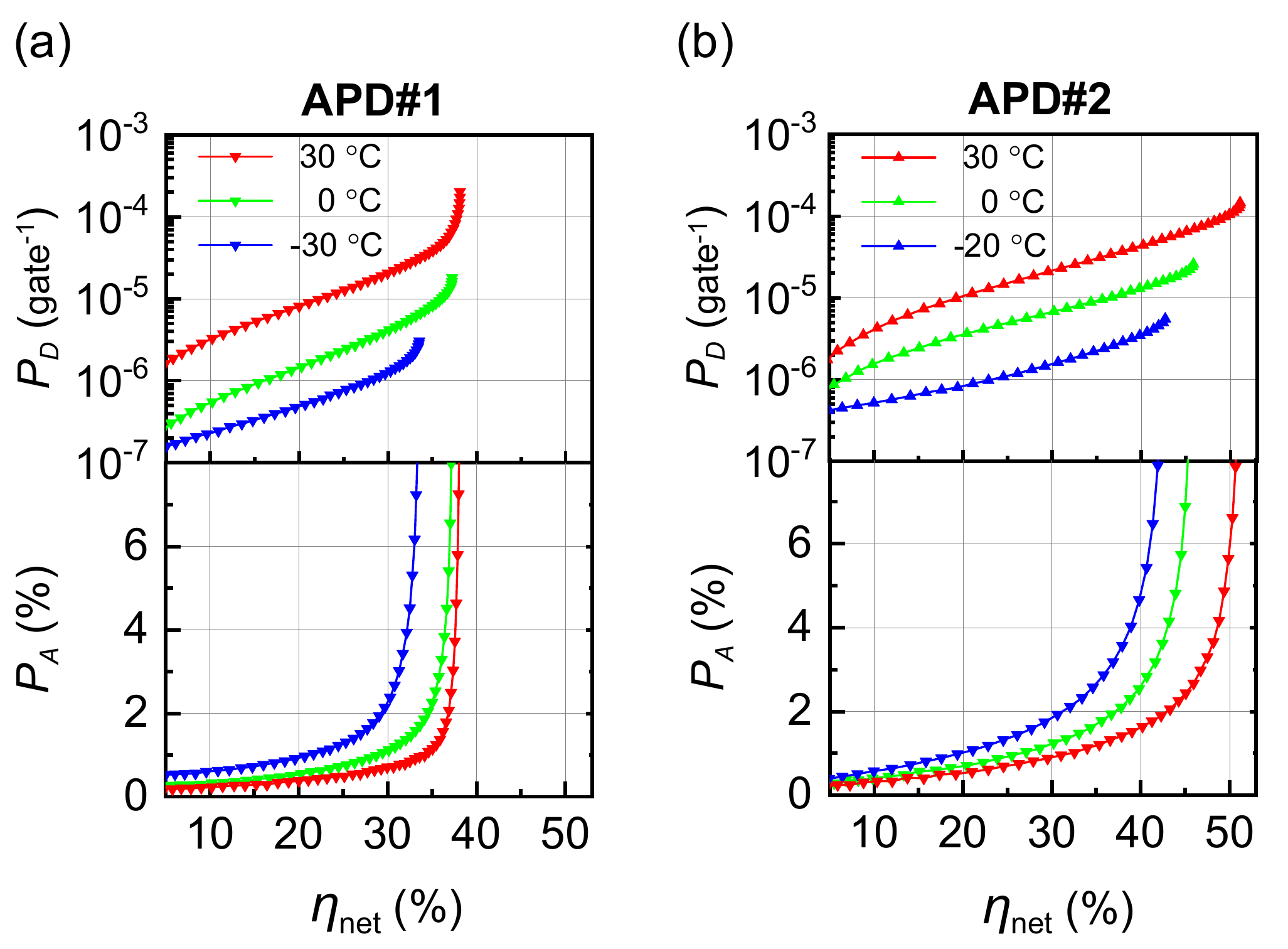}
\caption{Dark count (top) and afterpulse (bottom) 
 probabilities as a function of photon detection efficiency of \textbf{(a)} APD\#1  and \textbf{(b)} APD\#2 measured for several different temperatures.
}
\label{fig:003}
\newpage
\end{figure}

\begin{figure}[t]
\centering
\includegraphics[width=.8\linewidth]{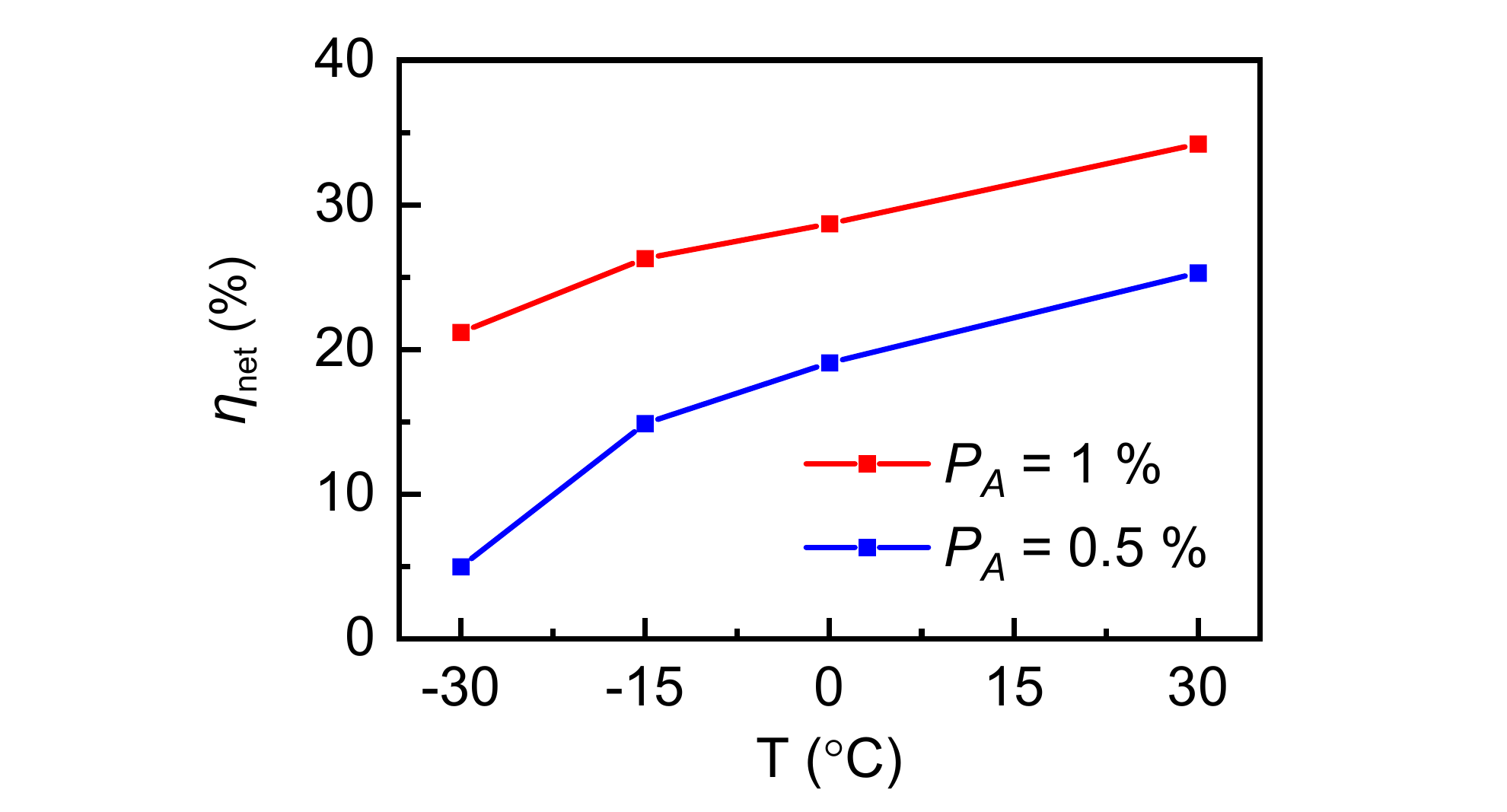}
\caption{Temperature dependencies of the photon detection efficiency for APD\#1 at the given afterpulsing probabilities of 0.5 \% (blue) and 1 \% (red).}
\label{fig:004}
\newpage
\end{figure}

Figure~\ref{fig:003} shows the characterisation results for APD\#1 and APD\#2.
We fixed the amplitude of the 1.25~GHz sinusoidal signal at 27.0~V, and measured the relevant parameters as a function of the applied direct current (DC) bias, but for clarity the results are plotted as a function of the net detection efficiency ($\eta_{net}$). Each device was measured at several different temperatures, while APD\#2 could only be cooled to -20~$^\circ$C due to the incompatibility of its thermal-electric cooler with the temperature control driver.  
Qualitatively, two devices behave similarly. 
Both dark count and afterpulsing probabilities increase with photon detection efficiency, and exhibit opposite dependencies on temperature.  For both APDs at $\eta_\mathrm{net} = 30~\%$, the afterpulsing probabilities are less than 2.3~\% at their lowest measurement temperatures with corresponding dark count probabilities of $1.25 \times 10^{-6}$ and $1.6 \times 10^{-6}$ for APD\#1 (-30~$^\circ$C) and APD\#2 (-20~$^\circ$C), respectively.   
Moreover, our UNIC-APDs can offer record low afterpulsing probabilities, as summarised for APD\#1 in Figure~\ref{fig:004}.  At -30~$^\circ$C, APD\#1 is able to achieve 5~\% and 21.2~\% detection efficiencies at 0.5~\% and 1.0~\% afterpulsing probabilities.  At these afterpulsing probabilities, the maximum detection efficiency increases with temperature and reaches 25.3~\% and 34.2~\% at 30~$^\circ$C. At a raised afterpulsing probability of 5.9~\%,  APD\#2 reaches a detection efficiency of 50~\% at a dark count probability of $1.1 \times 10^{-4}$ and a temperature of 30~$^\circ$C.

\begin{figure}[b]
\centering
\includegraphics[width=.8\linewidth]{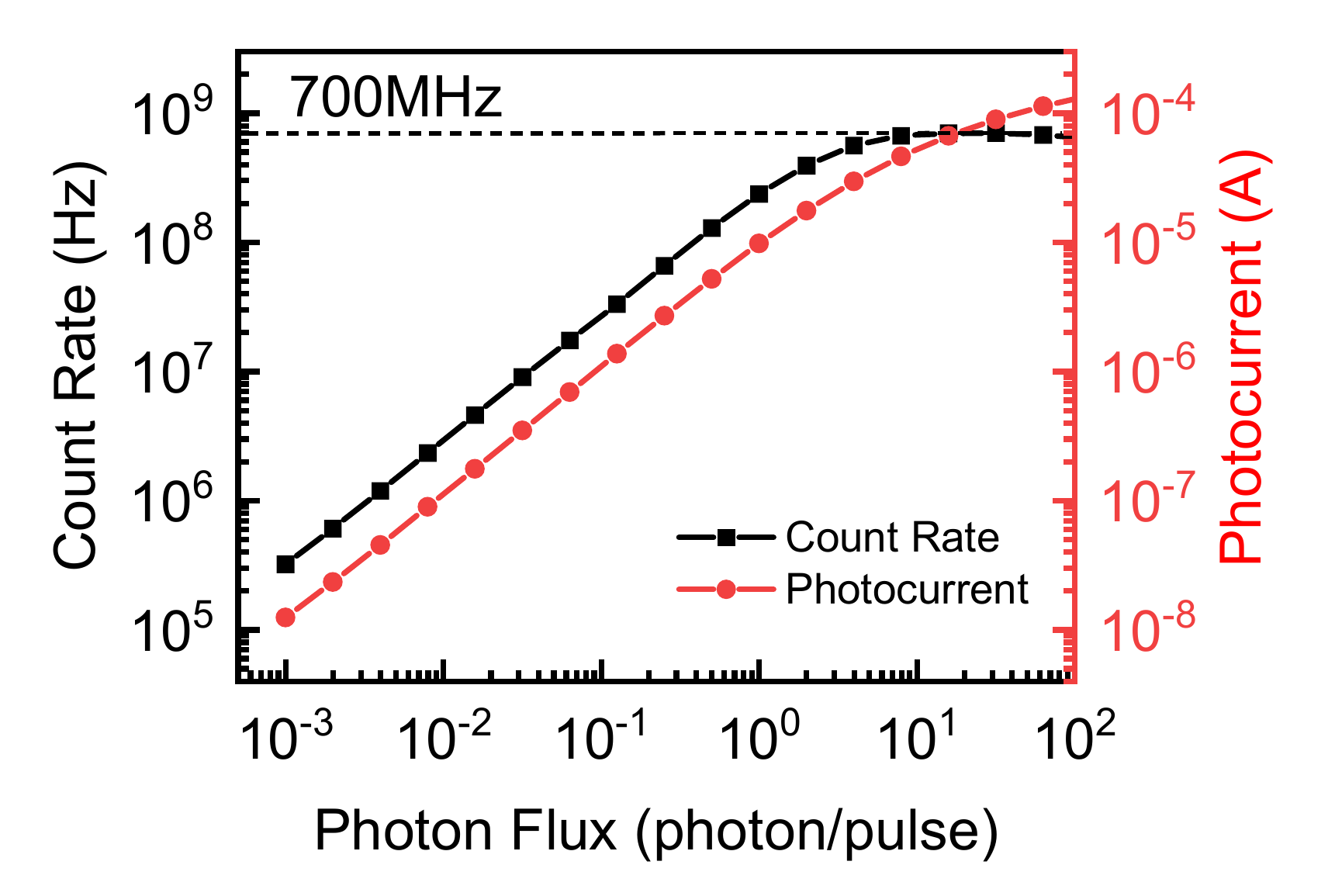}
\caption{Maximum count rate (blue) and photocurrent (red) \textit{vs } incident flux  for APD\#1.
}
\label{fig:005}
\newpage
\end{figure}

Maximum count rate is a crucial parameter for a number of applications, for example, high bit rate QKD \cite{yuan2018} and rapid phase tracking in twin-field QKD \cite{lucamarini18,zhou22}.  
To determine their maximum count rates, we used a DFB laser transmitting at 1.25~GHz as the illumination source and measure the count rate as a function of photon flux. 
Figure~\ref{fig:005} shows an exemplar result obtained from APD\#1 at a temperature of 
30~$^\circ$C with its detection efficiency set to 25.3~\% in the low flux regime.  The detector maintains a linear dependence with incident flux for count rates exceeding  100~MHz, while a maximum count rate of 700~MHz is obtained at the few photons/pulse regime.
We attribute the high count rate to the UNIC's ability of removing the capacitive response and thus allowing discrimination of faint avalanches.  From the accompanying current measurement, we extract an average avalanche charge of 38~fC, comparable to the best value of 35~fC \cite{yuan2010} obtained with the identical photocurrent measurement method. The ability to detect such weak avalanches ensures low afterpulsing probabilities in our UNIC-APDs.
APD\#2 was measured to have a similar avalanche charge as that of APD\#1.  When setting its  efficiency to 50~\%,  APD\#2's avalanche charge rose to 65~fC due to stronger bias applied. Nevertheless, it was still able to achieve a maximum count rate of 600~MHz.

\begin{table*}[h]
		\caption{\label{tab:Comparison} Performance comparison of sub-nanosecond gated  InGaAs detectors using different types of readout circuits.}
        \setlength{\tabcolsep}{0.4mm}{
	\centering
	\begin{tabular}{ccccccc}
		\hline\hline	
		& $ P_{\mathrm{A}}$(\%)& $\eta_\mathrm{net}$ (\%)&$P _{\mathrm{D}}$ ($ \mathrm{gate}^{\mathrm{-1}}$)&T ($^{\circ}$C) &   $f_g$ (GHz)& Readout Method
		\\
		\hline	\hline
  		
		This work &  1.0  &21.2 & 5.4$\times10^{-7} $  & -30 &1.25 &	UNIC \\
        \hline
		
		He \textit{et al}\cite{He2017} &1.0  &20.7 & 7.6$\times10^{-7}$     &-30 &1.00  &low-pass filter +\\ & & & & & & variable width discriminator  \\
		\hline
		
		Tada \textit{et al}\cite{Tada2020} &1.8 &27.7 & 8$\times10^{-7}$   &-35 &1.27 
        & band stop filter		\\
		\hline
  
		Fang \textit{et al}\cite{Fang2020}  &2.5 &20 & 1.1$\times10^{-6}$  &-30 &1.25 &low-pass filter  \\
        \hline

		Comandar \textit{et al}\cite{Comandar2015} &2.9 &20 &  1.0$\times10^{-6}$ &-30 &1.00 &self-differencing 	\\
		\hline
  
  	Liang \textit{et al}\cite{Liang2019} &4.5 &20 &  3.2$\times10^{-6}$   &-30          &1.25  &reference subtraction \\

   	\hline\hline

	\end{tabular}}
	
\end{table*}

Table~\ref{tab:Comparison} compares our results with those gigahertz-gated detectors equipped with different readout circuits. For impartiality,  we list just data measured at a fixed temperature of -30~$^\circ$C whenever possible.  Here, our UNIC-APD achieved an impressive 1\% afterpulsing probability at $\eta_\mathrm{net} =21.2$~\%, considerably outperforming most other methods among filtering\cite{Fang2020,Tada2020}, self-differencing \cite{Comandar2015} and reference subtraction \cite{Liang2019}.  
In terms of detection efficiency, our result improves marginally over the previous best\cite{He2017}, but which was achieved with help of a custom variable width discriminator to mitigate signal distortion by excessive filtering.  We attribute the outstanding performance of our detectors to low-distortion signal processing by UNIC's. 

It is useful to compare our UNIC-APDs with detectors deployed in QKD systems.  In the QKD system optimised for secure key rates (SKRs) \cite{yuan2018},  the room-temperature self-differencing detectors featured $f_g = 1$~GHz,  $\eta_\mathrm{net} = 31$ \%, $P_A = 4.4$\% and $P_D = 2.25 \times 10^{-4}$ and a SKR of 13.72~Mb/s over a 2~dB channel was obtained.  Our UNIC-APD could outperform in all these parameters. At 30~$^\circ$C and with $P_A =4.4$~\%, APD\#2 offers a higher efficiency of 49~\% efficiency and twice lower dark count probability of $9.4 \times 10^{-5}$, see Fig.~\ref{fig:003}\textbf{b}.  Combined with its high count capability, UNIC detectors are expected to allow a SKR exceeding 25~Mb/s over the same channel loss.  This provides an interesting technological path towards 100~Mb/s QKD via wavelength multiplexing. 

\section*{Conclusion}
To summarise, we have developed a novel approach of using UNICs for reading out avalanche signals from 1.25~GHz sinusoidally gated InGaAs APDs. 
UNIC-APDs were characterised to exhibit excellent performance across the temperature range between $\pm30^\circ$C, and can offer \textgreater20~\% detection efficiency at an ultra low afterpulsing probability of 1~\%.   This performance, together with the circuit's compactness and manufacturing tolerance, will allow UNIC-APDs a considerable potential in QKD applications. 

\begin{backmatter}
\bmsection{Funding}
National Natural Science Foundation of China (62250710162).

\bmsection{Disclosures}
The authors declare that there are no conflicts of interest related to this article.

\bmsection{Data availability}
Data underlying the results presented in this paper are not publicly available at this time but may be obtained from the authors upon reasonable request. 
\end{backmatter}
%\appendix
%%%%%%%%%%%%%%%%%%%%%%% References %%%%%%%%%%%%%%%%%%%%%%%%%

%\bibliography{Reference}% Produces the bibliography via BibTeX.

%\newpage

\end{document}